\documentclass[aps,prl,twocolumn,showpacs,amsfonts,amssymb,amsmath,letterpaper]{revtex4}
\usepackage[dvips]{graphicx}
\newcommand {\be}{\begin{equation}}
\newcommand {\ee}{\end{equation}}

\begin{document}

\title{Adaptive Resolution Simulation of Liquid Water}
\author{Matej Praprotnik}
\altaffiliation{On leave from the National Institute of Chemistry, Hajdrihova 19,
                 SI-1001 Ljubljana, Slovenia. Electronic Mail: praprot@cmm.ki.si}
\author{Luigi Delle Site}
\author{Kurt Kremer}
\affiliation{Max-Planck-Institut f\"ur Polymerforschung, Ackermannweg 10, D-55128 Mainz, Germany}
\author{Silvina Matysiak}
\author{Cecilia Clementi}
\affiliation{Department of Chemistry, Rice University, 6100 Main Street, Houston, Texas 77005}

\date\today
\begin{abstract}
We present a multiscale simulation of liquid water where a spatially adaptive
molecular resolution procedure allows for changing on-the-fly from a
coarse-grained to an all-atom representation. We show that this approach leads
to the correct description of all essential thermodynamic and structural
properties of liquid water.
\end{abstract}

\maketitle

Scientific problems in physics are usually classified by typical
scales of measurable quantities such as energy, length or time.
The appropriately determined scale together with a tailored model
and/or experiment is at the outset of a successful study. The
specific properties under investigation then typically define the
appropriate level of resolution. Using the optimal set of degrees
of freedom (DOFs) guarantees efficiency, accuracy and avoids huge
amounts of unnecessary detail, which might even obscure the
underlying physics. This approach dates back to the very beginning
of modern physics and finds its logical continuation
in systematic coarse-graining efforts
for modern computational materials science and biophysics
problems~\cite{Nielsen:2004,Izvekov:2006,Harmandaris:2006,Das:2005},
where full blown all-atom simulations are often beyond the
possibilities of current and near future computers. However, in many
problems of materials science and biology different time- and
length-scales are intrinsically interconnected, far beyond giving
constant prefactors on the next coarser level of detail.
Multiscale simulations are emerging as a promising tool for such
problems~\cite{DelleSite:2002,Neri:2005,Fabritiis:2006,Voth2006,Gunsterenjcp2006}.
Ideally, in addition to coarse-graining, one would like to make the step back to
characteristic fine grained configurations as well~\cite{Hess:2006,HeathKavrakiClementi2006}.
Recently, an adaptive multiscale scheme (AdResS) has been proposed by some of
us that also allows a free exchange of particles between regions
of different resolutions and to adjust locally the level of detail
at will, while maintaining equilibrium with the fluctuating
environment~\cite{Praprotnik}.

\begin{figure}[t!]
\centering
\includegraphics[width=1.0\linewidth]{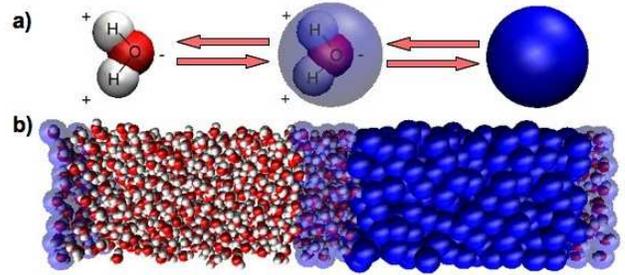}
\caption{\footnotesize (Color) On-the-fly interchange between the
all-atom and coarse-grained water models. a) The explicit all-atom
water molecule is represented at the left, and the coarse-grained
molecule at the right. The middle hybrid molecule interpolates
between the two, cf. text. A schematic representation of the full
system is shown in b), where a hybrid region connects the explicit
and coarse-grained levels of description. All the results
presented in the paper were obtained by performing NVT simulations
using ESPResSo~\cite{espresso} with a Langevin thermostat, with a
friction constant $\Gamma=5ps^{-1}$ and a time step of $0.002 ps$
at $T_{ref}=300K$ and $\rho=0.96 g/cm^{3}$ (the density was
obtained from an NPT simulation with $P_{ref}=1 \;atm$). Periodic
boundary conditions were applied in all directions. The box size
is 94.5 \AA \ in the x and 22 \AA \ in the y and z direction. The
width of the interface layer is 18.9 \AA \ in the x direction.
\label{fig.mix} } \vspace{-0.6cm}
\end{figure}
In this letter we address the above issue for liquid water, which
continues to be a very active research field for the obvious
reason, that all life exists in a water environment. Although many
classical models have been developed for the simulation of liquid
water~\cite{Jorgensen,Jorgensen1,Praprotnik:2005:2, Guillot:2002},
there is still no unique model that can reproduce all its
anomalous properties. Water is intrinsically multiscale, as it
plays different roles at different scales. For instance, in the
case of biomolecules in solution, at short length scales the
hydrogen bonding with water governs the local shape and stability of
folded biopolymers, whereas the ``hydrophobic effect" drives the organization
of the system at longer time and length scales (as for instance
the formation of membranes in aqueous solution).
The approach presented in this letter represent
a first step towards multiscale modeling of such complex
scenarios. For fluctuating systems, such as adsorption of water on
a hydrophilic surface, we envision a system in which an atomistic
description is needed only near the surface, and a mesoscopic
resolution can be used for water molecules further away from it,
with molecules freely moving around without ''feeling'' the local
level of resolution.

To accomplish this goal, we first propose a new single-site water model that
reproduces remarkably well the essential thermodynamic and structural features
of water, as obtained by detailed all-atom simulations. In the second step we
adapt the AdResS scheme to define a robust and physically accurate procedure to
smoothly join the explicit atomistic and coarse-grained resolution regions. The
resulting multiscale hybrid/mesoscopic model
system~\cite{Praprotnik,Praprotnik1,Praprotnik2} is composed of explicit
and coarse-grained molecules as presented in Figure~\ref{fig.mix}.
%
\paragraph{\textbf{Coarse-graining of water --}}
Recent work has focused on simplified coarse-grained models that
can reproduce qualitatively the all-atom center-of-mass (cm)
radial distribution function (rdf) of
water~\cite{Ashbaugh,Vothjcp2005,Gordon,Soper,Nezbeda}. To reproduce
further structural properties, such as the orientational
preferences for nearest neighbor configurations (i.e. for hydrogen
bonding), previous one-site models~\cite{Vothjcp2005,Gordon}
require a posteriori adjustment of the local
environment~\cite{Gordon}. The one-site model of water presented in
this letter can reproduce the structural and thermodynamic
properties of a widely used all-atom water model (namely, Rigid
TIP3P~\cite{Jorgensen}) with remarkable accuracy, without requiring any
orientational corrections~\cite{water3}. To construct the model we follow an
iterative inverse statistical mechanics approach proposed by Lyubarstev
\textit{et al.}~\cite{Lyubarstev} (for alternative procedures see also
Refs.~\cite{Izvekov:2006} and~\cite{Reith}). The aim of this scheme is
to numerically build a coarse-grained effective Hamiltonian that
can mimic the behavior of a set of physical observables of the
system under consideration (see i.e. Matysiak \textit{et
al.}~\cite{Matysiak_JMB2004,Matysiak_JMB2006} where this idea
has been used to define a coarse-grained protein Hamiltonian
''anchored'' to experimental data).
Additionally, in order to match the pressure of the coarse-grained to the
all-atom model, after each iteration, a weak constant force is
added to the effective force in such a way that the total
effective force and potential energy are zero at the $7\AA$ cutoff
distance~\cite{Reith, Praprotnik1}.
%
%
\begin{figure}[t!]
\centering
\includegraphics[width=0.9\linewidth]{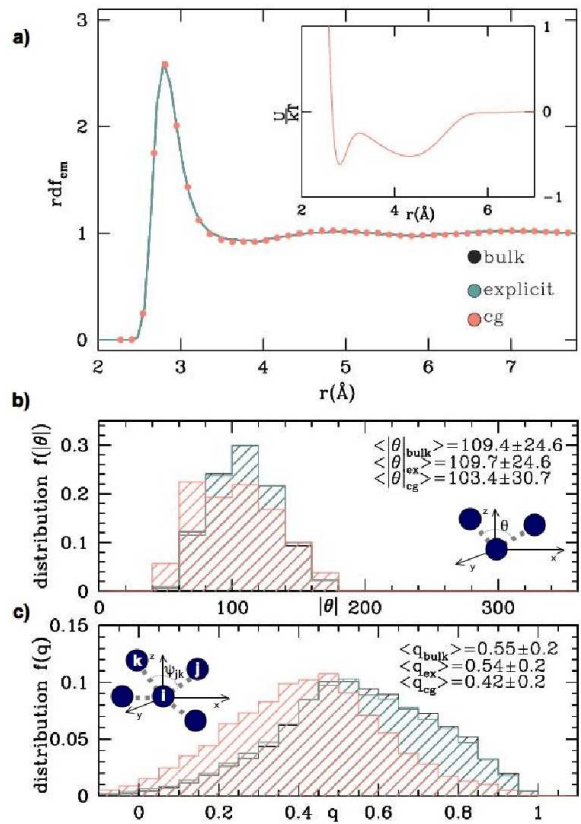}
\caption{\scriptsize (Color) a) The center-of-mass rdfs
for explicit (ex) [cyan line] and coarse-grained (cg) [red line] regions of the
hybrid system, are shown together with the rdfs corresponding to a bulk all-atom
simulation (black line). The local O-H and H-H rdfs for the
explicit molecules in the hybrid system compare
equally well to the standard bulk simulations (not shown).
The optimized effective potential for the coarse-grained model is shown in the
inset as a function of inter-particle separation, $r$.
(b) The center-of-mass angular distribution between three nearest
neighbors for the three cases studied as given in (a).
(c) The analogous distributions of the orientational order parameter $q$.
\label{fig.struct} }
\vspace{-0.6cm}
\end{figure}

Figure~\ref{fig.struct} shows the final results obtained when this
procedure is applied to a box of water molecules. A perfect
agreement between the all-atom and coarse-grained center-of-mass
rdfs is reached with our optimized effective potential (shown in
the inset of the figure) after 8 iterations. The effective potential has a
first primary minimum at about 2.8 \AA \; corresponding to the
first peak in the center-of-mass rdf. The slightly weaker
and significantly broader minimum at 4.5 \AA \ corresponds to the
second hydration shell. The combined effect of the two minima leads to
a local packing close to that of the all-atom TIP3P water. Our
effective coarse-grained potential is quite different from the
previously suggested potentials~\cite{Ashbaugh,Vothjcp2005,Gordon}:
while in previous one-site models the deepest minimum
corresponds to the second hydration shell, the absolute minimum in
our model is found in the first shell.
In order to more thoroughly quantify the structural properties of our
model (that are not completely defined by the rdf) we computed the angular
distribution between the center-of-mass of three nearest neighbor molecules,
and the distribution of the orientational order parameter $q$
as defined by Errington \textit{et al.}~\cite{Debenedetti}:
$q=1-\frac{3}{8}\sum_{j=1}^{3}\sum_{k=j+1}^{4}\biggl(cos\psi_{jk}+\frac{1}{3}\biggr)^2$,
where $\psi_{jk}$ is the angle formed by the lines
joining the oxygen atom of a given molecule and those of its nearest
neighbors $j$ and $k$. The parameter $q$ measures the extent to
which a molecule and its four nearest neighbors adopt a
tetrahedral arrangement~\cite{Debenedetti}.
The good agreement between the explicit and coarse-grained water models shown
in Figures~\ref{fig.struct} (b) and (c) indicates that although our
coarse-grained model is spherically symmetric and therefore does not
have any explicit directionality, it effectively captures the orientational
preferences of hydrogen bonding.

As a consequence of the reduced number of DOFs, there is a time
scale difference in the dynamics of the coarse-grained system,
that is faster than what is predicted from all-atom simulations.
The diffusion coefficient of the center-of-mass for the Rigid
TIP3P model is $D \approx (3.7 \pm 0.3) \times 10 ^{-9}
\frac{m^2}{s}$ while for the coarse-grained model is $D \approx
(8.1 \pm 0.4) \times 10^{-9} \frac{m^2}{s}$. While an accelerated
time scale can be very advantageous in some cases, one can also
adjust $D$ by an increased background friction in the Langevin
thermostat~\cite{Kremer:1990,Izvekov:2006}. \noindent
\paragraph{\textbf{Multiscale procedure --}}
In order to smoothly join the atomistic and coarse-grained
resolutions, we apply the AdResS scheme of Praprotnik \textit{et
al.}~\cite{Praprotnik,Praprotnik1} to couple a model system
composed of explicit and coarse-grained molecules. Half of the simulation box
is occupied by atomistic (Rigid TIP3P) water molecules while the other half is
filled with the same number of corresponding coarse-grained molecules as
schematically presented in Figure~\ref{fig.mix}. The two regions freely
exchange molecules through a
transition regime, where the molecules change their resolution and
their number of DOFs accordingly. The interface region contains
hybrid molecules that are composed of an all-atom molecule
with an additional massless center-of-mass particle serving as an
interaction site. The transition, which needs to be smooth in order to be
used in MD simulations, is governed by a weighting function
$w(x)\in[0,1]$ that interpolates  the interaction forces between the two
regimes, and assigns the identity of the particle.
We used the weighting function defined in~\cite{Praprotnik}, in such a way that
$w=1$ corresponds to the atomistic region, and $w=0$ to the coarse-grained
region, whereas the values $0<w<1$ correspond to the interface layer. The atomic
and mesoscopic length scales are coupled via the intermolecular
force acting between centers of mass of molecule $\alpha$ and
$\beta$ as:
\be
{\bf F}_{\alpha\beta}=w(X_{\alpha})w(X_{\beta}){\bf F}^{atom}_{\alpha\beta} +
[1-w(X_{\alpha})w(X_{\beta})]{\bf F}^{cm}_{\alpha\beta},\label{eq:AdResS}
\ee
where
${\bf F}^{atom}_{\alpha\beta}=\sum_{i_{\alpha}, j_{\beta}} {\bf
F}^{atom}_{i_{\alpha} j_{\beta}}= -\sum_{i_{\alpha},
j_{\beta}}\frac{\partial U^{atom}}{\partial {\bf r}_{i_{\alpha}
j_{\beta}}}$
is the sum of all pair intermolecular atom
interactions between explicit atoms of the molecules $\alpha$ and
$\beta$ and ${\bf F}^{cm}_{\alpha\beta}=-\frac{\partial
  U^{cm}}{\partial {\bf R}_{\alpha\beta}}$ is the corresponding
effective intermolecular force between their centers of mass.
${\bf r}_{i_{\alpha} j_{\beta}}= {\bf r}_{i_{\alpha}}-{\bf
r}_{j_{\beta}}$ is the vector between atom $i$ in molecule
$\alpha$ and atom $j$ in molecule $\beta$ and ${\bf R}_{\alpha
\beta}= {\bf R}_{\alpha}-{\bf R}_{\beta}$ the vector between the
centers of mass of molecules $\alpha$ and $\beta$, with the
corresponding $X$ coordinates $X_{\alpha}$ and $X_{\beta}$. Each
time a molecule crosses a boundary between the different regimes
it gains or looses (depending on whether it leaves or enters the
coarse-grained region) its equilibrated rotational DOFs while
retaining its linear momentum. To supply or  remove the latent
heat caused by the switch of resolution this method is employed
together with a Langevin thermostat~\cite{Praprotnik}. As
discussed in an earlier publication, it is important to
interpolate the forces and not the interaction potential if the
Newton's Third Law is to be satisfied~\cite{Praprotnik2}.
The key methodological issue in applying our approach to water is
how to treat the long-range electrostatic interactions. We have
chosen the reaction field (RF) method, in which all molecules with
the charge center outside a spherical cavity of a molecular based
cutoff radius $R_c=9 \AA$ are treated as a dielectric continuum
with a dielectric constant $\epsilon_{RF}$~\cite{Neumann:1983,
Neumann:1985, Tironi, Praprotnik:2004}. The Coulomb force acting
on a charge $e_{i_{\alpha}}$, at the center of the cutoff sphere,
due to a charge $e_{j_{\beta}}$ within the cavity is:
\be
{\bf F}^{atom}_{C_{i_{\alpha}
j_{\beta}}}({\bf r}_{i_{\alpha} j_{\beta}})=
\frac{e_{i_{\alpha}}e_{j_{\beta}}}{4 \pi\epsilon_0}\biggl
[\frac{1}{r_{i_{\alpha}
j_{\beta}}^3}-\frac{1}{R_c^3}\frac{2(\epsilon_{RF}-1)}{1+2\epsilon_{RF}}
\biggr]{\bf r}_{i_{\alpha} j_{\beta}}.
\label{eq:rf}
\ee
There are three main reasons that make the RF method a natural
choice to be applied with our scheme (Eq.~\ref{eq:AdResS}):
the first reason is that it does not impose any artificial periodicity
(that would interfere with the definition of the interface layer of
the system), the second one is the
pairwise form of the reaction field term (Eq.~\ref{eq:rf}),
and the third one is that, as our scheme, the RF method works
accurately only in a combination with a thermostat~\cite{Tironi}.
The last point can be better understood if the switching of the resolution
is seen as a geometry induced phase transition~\cite{Praprotnik2}.
Detailed comparisons between
the bulk explicit simulations and the explicit regime in our
hybrid setup prove that this approach does not alter the
structural properties of the water model studied.
The width of the interface layer is determined by the maximum
range of electrostatic interactions, in order to allow the
molecules to equilibrate on-the-fly while moving from one resolution to the
other.
In order to obtain a very smooth density distribution in the interface layer we use
a conservative choice, that is, double the maximum range of interactions; however,
a smaller interface layer can be used without affecting the fluxes between the
layers, nor the results in the two resolution regimes.
We observed that for a system containing only hybrid molecules, the pressure is
increased in comparison to the reference all-atom system with the maximum
deviation at $w=1/2$ ($280\%$ difference). This problem is solved by employing
an interface pressure correction~\cite{Praprotnik1}.

\begin{figure}[t!]
\centering
\includegraphics[height=0.8\linewidth]{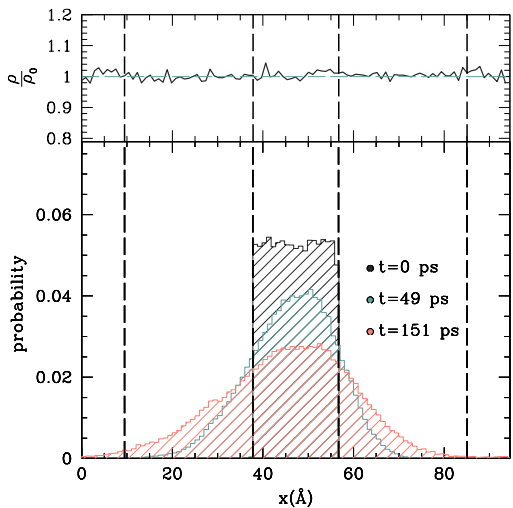}
\caption{\scriptsize
(Color) Top figure: normalized density profile in the x-direction of the hybrid system.
Bottom figure: Time evolution of  a diffusion profile for the
molecules that are initially (at time $t=0$ ps) located in the interface region.
The diffusion profile is averaged over $\approx 400$ different time origins.
Vertical lines denote the boundaries of the interface layer.}
\label{fig.diff}
\vspace{-0.6cm}
\end{figure}

The structural (see Figure~\ref{fig.struct}) and thermodynamic
properties of the bulk all-atom simulations are correctly
reproduced in the hybrid system with the same mean temperature
($0.1\%$ difference) and pressure ($0.5\%$ difference). The
normalized density for the hybrid system is homogeneous in the
coarse-grained and explicit regions with (very) small oscillations
in the transition regime, cf. Figure~\ref{fig.diff}.
In order to prove the free exchange of molecules between the different
regimes we have computed the time evolution of a diffusion profile for
molecules that were initially localized at the interface layer.
Figure~\ref{fig.diff} shows that these molecules spread out asymmetrically with
time. This asymmetry arises from the aforementioned difference in diffusion
coefficient between the all-atom and coarse-grained regions.
It is possible to obtain the same diffusional dynamics across the
different resolutions by adding a local Langevin thermostat, as
mentioned above~\cite{water2}.
However, it is worth noticing that this difference in time scale
can be advantageous for reaching longer simulation times in
systems where multiple length and time scales are intrinsically
present.

In this letter a general and computationally efficient multiscale
model for water has been presented. We have shown that a one-site
coarse-grained model for water molecules can reproduce remarkably
well thermodynamic and structural properties of the Rigid TIP3P water
model, and that a smooth transition and free exchange between
coarse-grained and all-atom resolutions is possible.
We stress that the presented multiscale approach can be applied to
any other either flexible or rigid nonporalizable classical water
model, e.g. SPC or
SPC/E~\cite{Berendsen:1981,Berendsen:1987,Praprotnik:2004}. We
envision that such a multiscale resolution of water will play an
important role in the modeling of wet/dry interfaces and
biomolecular simulations.

\begin{acknowledgments}
We wish to thank the NSF-funded Institute for Pure and Applied
Mathematics at UCLA where this work was first planned. This work
has been supported in part by grants from NSF, Texas-ATP, the
Robert A. Welch Foundation (C.C.) and the Volkswagen foundation
(K.K \& L.~D.~S.). The  Rice University Cray XD1 Cluster ADA
used for the calculations is supported by NSF, Intel, and Hewlett Packard.
\end{acknowledgments}


\begin{thebibliography}{35}
\expandafter\ifx\csname natexlab\endcsname\relax\def\natexlab#1{#1}\fi
\expandafter\ifx\csname bibnamefont\endcsname\relax
  \def\bibnamefont#1{#1}\fi
\expandafter\ifx\csname bibfnamefont\endcsname\relax
  \def\bibfnamefont#1{#1}\fi
\expandafter\ifx\csname citenamefont\endcsname\relax
  \def\citenamefont#1{#1}\fi
\expandafter\ifx\csname url\endcsname\relax
  \def\url#1{\texttt{#1}}\fi
\expandafter\ifx\csname urlprefix\endcsname\relax\def\urlprefix{URL }\fi
\providecommand{\bibinfo}[2]{#2}
\providecommand{\eprint}[2][]{\url{#2}}

\bibitem[{\citenamefont{Nielsen et~al.}(2004)\citenamefont{Nielsen, Lopez,
  Srinivas, and Klein}}]{Nielsen:2004}
\bibinfo{author}{\bibfnamefont{S.~O.} \bibnamefont{Nielsen}},
  \bibinfo{author}{\bibfnamefont{C.~F.} \bibnamefont{Lopez}},
  \bibinfo{author}{\bibfnamefont{G.}~\bibnamefont{Srinivas}}, \bibnamefont{and}
  \bibinfo{author}{\bibfnamefont{M.~L.} \bibnamefont{Klein}},
  \bibinfo{journal}{J. Phys.: Condens. Matter} \textbf{\bibinfo{volume}{16}},
  \bibinfo{pages}{R481} (\bibinfo{year}{2004}).


\bibitem[{\citenamefont{Izvekov and Voth}(2006)}]{Izvekov:2006}
\bibinfo{author}{\bibfnamefont{S.}~\bibnamefont{Izvekov}} \bibnamefont{and}
  \bibinfo{author}{\bibfnamefont{G.~A.} \bibnamefont{Voth}},
  \bibinfo{journal}{J. Chem. Phys.} \textbf{\bibinfo{volume}{125}},
  \bibinfo{pages}{151101} (\bibinfo{year}{2006}).

\bibitem[{\citenamefont{Harmandaris et~al.}(2006)\citenamefont{Harmandaris,
  Adhikari, Van~der Vegt, and Kremer}}]{Harmandaris:2006}
\bibinfo{author}{\bibfnamefont{V.~A.} \bibnamefont{Harmandaris}},
  \bibinfo{author}{\bibfnamefont{N.~P.} \bibnamefont{Adhikari}},
  \bibinfo{author}{\bibfnamefont{N.~F.~A.} \bibnamefont{Van~der Vegt}},
  \bibnamefont{and} \bibinfo{author}{\bibfnamefont{K.}~\bibnamefont{Kremer}},
  \bibinfo{journal}{Macromolecules} \textbf{\bibinfo{volume}{39}},
  \bibinfo{pages}{6708} (\bibinfo{year}{2006}).

\bibitem[{\citenamefont{Das et~al.}(2005)\citenamefont{Das,
  Matysiak, and Clementi}}]{Das:2005}
\bibinfo{author}{\bibfnamefont{P.} \bibnamefont{Das}},
  \bibinfo{author}{\bibfnamefont{S.} \bibnamefont{Matysiak}},
  \bibnamefont{and} \bibinfo{author}{\bibfnamefont{C.}~\bibnamefont{Clementi}},
  \bibinfo{journal}{Proc. Natl. Acad. Sci. USA} \textbf{\bibinfo{volume}{102}},
  \bibinfo{pages}{10141} (\bibinfo{year}{2005}).

\bibitem[{\citenamefont{{Delle Site} et~al.}(2002)\citenamefont{{Delle Site},
  Abrams, Alavi, and Kremer}}]{DelleSite:2002}
\bibinfo{author}{\bibfnamefont{L.}~\bibnamefont{{Delle Site}}},
  \bibinfo{author}{\bibfnamefont{C.~F.} \bibnamefont{Abrams}},
  \bibinfo{author}{\bibfnamefont{A.}~\bibnamefont{Alavi}}, \bibnamefont{and}
  \bibinfo{author}{\bibfnamefont{K.}~\bibnamefont{Kremer}},
  \bibinfo{journal}{Phys. Rev. Lett.} \textbf{\bibinfo{volume}{89}},
  \bibinfo{pages}{156103} (\bibinfo{year}{2002}).

\bibitem[{\citenamefont{Neri et~al.}(2005)\citenamefont{Neri, Anselmi,
  Cascella, Maritan, and Carloni}}]{Neri:2005}
\bibinfo{author}{\bibfnamefont{M.}~\bibnamefont{Neri}},
  \bibinfo{author}{\bibfnamefont{C.}~\bibnamefont{Anselmi}},
  \bibinfo{author}{\bibfnamefont{M.}~\bibnamefont{Cascella}},
  \bibinfo{author}{\bibfnamefont{A.}~\bibnamefont{Maritan}}, \bibnamefont{and}
  \bibinfo{author}{\bibfnamefont{P.}~\bibnamefont{Carloni}},
  \bibinfo{journal}{Phys. Rev. Lett.} \textbf{\bibinfo{volume}{95}},
  \bibinfo{pages}{218102} (\bibinfo{year}{2005}).

\bibitem[{\citenamefont{Fabritiis et~al.}(2006)\citenamefont{Fabritiis,
  Delgado-Buscalioni, and Coveney}}]{Fabritiis:2006}
\bibinfo{author}{\bibfnamefont{G.~D.} \bibnamefont{Fabritiis}},
  \bibinfo{author}{\bibfnamefont{R.}~\bibnamefont{Delgado-Buscalioni}},
  \bibnamefont{and} \bibinfo{author}{\bibfnamefont{P.~V.}
  \bibnamefont{Coveney}}, \bibinfo{journal}{Phys. Rev. Lett.}
  \textbf{\bibinfo{volume}{97}}, \bibinfo{pages}{134501}
  (\bibinfo{year}{2006}).

\bibitem[{\citenamefont{Shi et~al.}(2006)\citenamefont{Shi, Izvekov, and
  Voth}}]{Voth2006}
\bibinfo{author}{\bibfnamefont{Q.}~\bibnamefont{Shi}},
  \bibinfo{author}{\bibfnamefont{S.}~\bibnamefont{Izvekov}}, \bibnamefont{and}
  \bibinfo{author}{\bibfnamefont{G.~A.} \bibnamefont{Voth}},
  \bibinfo{journal}{J. Phys. Chem. B} \textbf{\bibinfo{volume}{110}},
  \bibinfo{pages}{15045} (\bibinfo{year}{2006}).

\bibitem[{\citenamefont{Christen and van Gunsteren}(2006)}]{Gunsterenjcp2006}
\bibinfo{author}{\bibfnamefont{M.}~\bibnamefont{Christen}} \bibnamefont{and}
  \bibinfo{author}{\bibfnamefont{W.~F.} \bibnamefont{van Gunsteren}},
  \bibinfo{journal}{J.\ Chem.\ Phys.} \textbf{\bibinfo{volume}{124}},
  \bibinfo{pages}{154106} (\bibinfo{year}{2006}).

\bibitem[{\citenamefont{Heath et~al.}(2006)\citenamefont{Heath, Kavraki, and
  Clementi}}]{HeathKavrakiClementi2006}
\bibinfo{author}{\bibfnamefont{A.}~\bibnamefont{Heath}},
  \bibinfo{author}{\bibfnamefont{L.}~\bibnamefont{Kavraki}}, \bibnamefont{and}
  \bibinfo{author}{\bibfnamefont{C.}~\bibnamefont{Clementi}},
  \bibinfo{journal}{Proteins: Struct. Funct. Bioinf.}
  \textbf{\bibinfo{volume}{in press}} (\bibinfo{year}{2006}).

\bibitem[{\citenamefont{Hess et~al.}(2006)\citenamefont{Hess, Leon, Van~der
  Vegt, and Kremer}}]{Hess:2006}
\bibinfo{author}{\bibfnamefont{B.}~\bibnamefont{Hess}},
  \bibinfo{author}{\bibfnamefont{S.}~\bibnamefont{Leon}},
  \bibinfo{author}{\bibfnamefont{N.}~\bibnamefont{Van~der Vegt}},
  \bibnamefont{and} \bibinfo{author}{\bibfnamefont{K.}~\bibnamefont{Kremer}},
  \bibinfo{journal}{Soft Matter} \textbf{\bibinfo{volume}{2}},
  \bibinfo{pages}{409} (\bibinfo{year}{2006}).

\bibitem[{\citenamefont{Praprotnik et~al.}(2005)\citenamefont{Praprotnik,
  Delle~Site, and Kremer}}]{Praprotnik}
\bibinfo{author}{\bibfnamefont{M.}~\bibnamefont{Praprotnik}},
  \bibinfo{author}{\bibfnamefont{L.}~\bibnamefont{Delle~Site}},
  \bibnamefont{and} \bibinfo{author}{\bibfnamefont{K.}~\bibnamefont{Kremer}},
  \bibinfo{journal}{J.\ Chem.\ Phys.} \textbf{\bibinfo{volume}{123}},
  \bibinfo{pages}{224106} (\bibinfo{year}{2005}).

\bibitem[{\citenamefont{Jorgensen et~al.}(1983)\citenamefont{Jorgensen,
  Chandrasekhar, Madura, Impey, and Klein}}]{Jorgensen}
\bibinfo{author}{\bibfnamefont{W.~L.} \bibnamefont{Jorgensen}},
  \bibinfo{author}{\bibfnamefont{J.}~\bibnamefont{Chandrasekhar}},
  \bibinfo{author}{\bibfnamefont{J.~D.} \bibnamefont{Madura}},
  \bibinfo{author}{\bibfnamefont{R.~W.} \bibnamefont{Impey}}, \bibnamefont{and}
  \bibinfo{author}{\bibfnamefont{M.~L.} \bibnamefont{Klein}},
  \bibinfo{journal}{J Chem Phys} \textbf{\bibinfo{volume}{79}},
  \bibinfo{pages}{926} (\bibinfo{year}{1983}).

\bibitem[{\citenamefont{Mahoney and Jorgensen}(2000)}]{Jorgensen1}
\bibinfo{author}{\bibfnamefont{M.~W.} \bibnamefont{Mahoney}} \bibnamefont{and}
  \bibinfo{author}{\bibfnamefont{W.~L.} \bibnamefont{Jorgensen}},
  \bibinfo{journal}{J.\ Chem.\ Phys.} \textbf{\bibinfo{volume}{112}},
  \bibinfo{pages}{8910} (\bibinfo{year}{2000}).


\bibitem[{\citenamefont{Praprotnik and
  Jane\v{z}i\v{c}}(2005)}]{Praprotnik:2005:2}
\bibinfo{author}{\bibfnamefont{M.}~\bibnamefont{Praprotnik}} \bibnamefont{and}
  \bibinfo{author}{\bibfnamefont{D.}~\bibnamefont{Jane\v{z}i\v{c}}},
  \bibinfo{journal}{J. Chem. Phys.} \textbf{\bibinfo{volume}{122}},
  \bibinfo{pages}{174103} (\bibinfo{year}{2005}).


\bibitem[{\citenamefont{Guillot}(2002)}]{Guillot:2002}
\bibinfo{author}{\bibfnamefont{B.}~\bibnamefont{Guillot}}, \bibinfo{journal}{J.
  Mol. Liq.} \textbf{\bibinfo{volume}{101}}, \bibinfo{pages}{219}
  (\bibinfo{year}{2002}).

\bibitem[{\citenamefont{Praprotnik
  et~al.}(2006{\natexlab{a}})\citenamefont{Praprotnik, Delle~Site, and
  Kremer}}]{Praprotnik1}
\bibinfo{author}{\bibfnamefont{M.}~\bibnamefont{Praprotnik}},
  \bibinfo{author}{\bibfnamefont{L.}~\bibnamefont{Delle~Site}},
  \bibnamefont{and} \bibinfo{author}{\bibfnamefont{K.}~\bibnamefont{Kremer}},
  \bibinfo{journal}{Phys.\ Rev.\ E} \textbf{\bibinfo{volume}{73}},
  \bibinfo{pages}{066701} (\bibinfo{year}{2006}{\natexlab{a}}).

\bibitem[{\citenamefont{Praprotnik
  et~al.}(2006{\natexlab{b}})\citenamefont{Praprotnik, Kremer, and {Delle
  Site}}}]{Praprotnik2}
\bibinfo{author}{\bibfnamefont{M.}~\bibnamefont{Praprotnik}},
  \bibinfo{author}{\bibfnamefont{K.}~\bibnamefont{Kremer}}, \bibnamefont{and}
  \bibinfo{author}{\bibfnamefont{L.}~\bibnamefont{{Delle Site}}},
  \bibinfo{journal}{Phys.\ Rev.\ E accepted, cond-mat/0609019}
  (\bibinfo{year}{2006}{\natexlab{b}}).

\bibitem[{\citenamefont{http://www.espresso.mpg.de.}()}]{espresso}
\bibinfo{author}{\bibnamefont{http://www.espresso.mpg.de.}}

\bibitem[{\citenamefont{Garde and Ashbaugh}(2001)}]{Ashbaugh}
\bibinfo{author}{\bibfnamefont{S.}~\bibnamefont{Garde}} \bibnamefont{and}
  \bibinfo{author}{\bibfnamefont{H.~S.} \bibnamefont{Ashbaugh}},
  \bibinfo{journal}{J.\ Chem.\ Phys.} \textbf{\bibinfo{volume}{115}},
  \bibinfo{pages}{977} (\bibinfo{year}{2001}).

\bibitem[{\citenamefont{Voth and Izvekov}(2005)}]{Vothjcp2005}
\bibinfo{author}{\bibfnamefont{G.~A.} \bibnamefont{Voth}} \bibnamefont{and}
  \bibinfo{author}{\bibfnamefont{S.}~\bibnamefont{Izvekov}},
  \bibinfo{journal}{J.\ Chem.\ Phys.} \textbf{\bibinfo{volume}{123}},
  \bibinfo{pages}{134105} (\bibinfo{year}{2005}).

\bibitem[{\citenamefont{Head-Gordon and Stillinger}(1993)}]{Gordon}
\bibinfo{author}{\bibfnamefont{T.}~\bibnamefont{Head-Gordon}} \bibnamefont{and}
  \bibinfo{author}{\bibfnamefont{F.~H.} \bibnamefont{Stillinger}},
  \bibinfo{journal}{J.\ Chem.\ Phys.} \textbf{\bibinfo{volume}{98}},
  \bibinfo{pages}{3313} (\bibinfo{year}{1993}).

\bibitem[{\citenamefont{Soper}(1996)}]{Soper}
\bibinfo{author}{\bibfnamefont{A.~K.} \bibnamefont{Soper}},
  \bibinfo{journal}{Chem Phys} \textbf{\bibinfo{volume}{202}},
  \bibinfo{pages}{295} (\bibinfo{year}{1996}).

\bibitem[{\citenamefont{Nezbeda}(2005)}]{Nezbeda}
\bibinfo{author}{\bibfnamefont{I.}~\bibnamefont{Nezbeda}},
  \bibinfo{journal}{Molec Phys} \textbf{\bibinfo{volume}{103}},
  \bibinfo{pages}{59} (\bibinfo{year}{2005}).

\bibitem[{wat()}]{water3}
\bibinfo{note}{Since a water molecule is electrically neutral the interaction
  site has a zero electric charge. Furthermore, in contrast to a Stockmayer
  fluid, the presented one-site model does not have a dipole (or any higher
  electric multipole). All the electrostatic interactions are therefore
  contained in the effective intermolecular potential.}

\bibitem[{\citenamefont{Lyubartsev and Laaksonen}(1995)}]{Lyubarstev}
\bibinfo{author}{\bibfnamefont{A.~P.} \bibnamefont{Lyubartsev}}
  \bibnamefont{and}
  \bibinfo{author}{\bibfnamefont{A.}~\bibnamefont{Laaksonen}},
  \bibinfo{journal}{Phys Rev E} \textbf{\bibinfo{volume}{52}},
  \bibinfo{pages}{3730} (\bibinfo{year}{1995}).

\bibitem[{\citenamefont{Reith et~al.}(2003)\citenamefont{Reith, Putz, and
  Muller-Plathe}}]{Reith}
\bibinfo{author}{\bibfnamefont{D.}~\bibnamefont{Reith}},
  \bibinfo{author}{\bibfnamefont{M.}~\bibnamefont{Putz}}, \bibnamefont{and}
  \bibinfo{author}{\bibfnamefont{F.}~\bibnamefont{Muller-Plathe}},
  \bibinfo{journal}{J Comput Chem} \textbf{\bibinfo{volume}{24}},
  \bibinfo{pages}{1624} (\bibinfo{year}{2003}).


\bibitem[{\citenamefont{Matysiak and Clementi}(2004)}]{Matysiak_JMB2004}
\bibinfo{author}{\bibfnamefont{S.}~\bibnamefont{Matysiak}} \bibnamefont{and}
  \bibinfo{author}{\bibfnamefont{C.}~\bibnamefont{Clementi}},
  \bibinfo{journal}{J. Mol. Biol.} \textbf{\bibinfo{volume}{343}},
  \bibinfo{pages}{235} (\bibinfo{year}{2004}).

\bibitem[{\citenamefont{Matysiak and Clementi}(2006)}]{Matysiak_JMB2006}
\bibinfo{author}{\bibfnamefont{S.}~\bibnamefont{Matysiak}} \bibnamefont{and}
  \bibinfo{author}{\bibfnamefont{C.}~\bibnamefont{Clementi}},
  \bibinfo{journal}{J. Mol. Biol.} \textbf{\bibinfo{volume}{363}},
  \bibinfo{pages}{297} (\bibinfo{year}{2006}).

\bibitem[{\citenamefont{Errington and Debenedetti}(2001)}]{Debenedetti}
\bibinfo{author}{\bibfnamefont{J.~R.} \bibnamefont{Errington}} \bibnamefont{and}
  \bibinfo{author}{\bibfnamefont{P.~G.} \bibnamefont{Debenedetti}},
  \bibinfo{journal}{Nature} \textbf{\bibinfo{volume}{409}},
  \bibinfo{pages}{318} (\bibinfo{year}{2001}).

\bibitem[{\citenamefont{Kremer and Grest}(1990)}]{Kremer:1990}
\bibinfo{author}{\bibfnamefont{K.}~\bibnamefont{Kremer}} \bibnamefont{and}
  \bibinfo{author}{\bibfnamefont{G.~S.} \bibnamefont{Grest}},
  \bibinfo{journal}{J. Chem. Phys.} \textbf{\bibinfo{volume}{92}},
  \bibinfo{pages}{5057} (\bibinfo{year}{1990}).

\bibitem[{\citenamefont{Neumann}(1983)}]{Neumann:1983}
\bibinfo{author}{\bibfnamefont{M.}~\bibnamefont{Neumann}},
  \bibinfo{journal}{Mol. Phys.} \textbf{\bibinfo{volume}{50}},
  \bibinfo{pages}{841} (\bibinfo{year}{1983}).

\bibitem[{\citenamefont{Neumann}(1985)}]{Neumann:1985}
\bibinfo{author}{\bibfnamefont{M.}~\bibnamefont{Neumann}}, \bibinfo{journal}{J.
  Chem. Phys.} \textbf{\bibinfo{volume}{82}}, \bibinfo{pages}{5663}
  (\bibinfo{year}{1985}).

\bibitem[{\citenamefont{Tironi et~al.}(1995)\citenamefont{Tironi, Sperb, Smith,
  and van Gunsteren}}]{Tironi}
\bibinfo{author}{\bibfnamefont{I.~G.} \bibnamefont{Tironi}},
  \bibinfo{author}{\bibfnamefont{R.}~\bibnamefont{Sperb}},
  \bibinfo{author}{\bibfnamefont{P.~E.} \bibnamefont{Smith}}, \bibnamefont{and}
  \bibinfo{author}{\bibfnamefont{W.~F.} \bibnamefont{van Gunsteren}},
  \bibinfo{journal}{J.\ Chem.\ Phys.} \textbf{\bibinfo{volume}{102}},
  \bibinfo{pages}{5451} (\bibinfo{year}{1995}).

\bibitem[{\citenamefont{Praprotnik et~al.}(2004)\citenamefont{Praprotnik,
  Jane\v{z}i\v{c}, and Mavri}}]{Praprotnik:2004}
\bibinfo{author}{\bibfnamefont{M.}~\bibnamefont{Praprotnik}},
  \bibinfo{author}{\bibfnamefont{D.}~\bibnamefont{Jane\v{z}i\v{c}}},
  \bibnamefont{and} \bibinfo{author}{\bibfnamefont{J.}~\bibnamefont{Mavri}},
  \bibinfo{journal}{J. Phys. Chem. A} \textbf{\bibinfo{volume}{108}},
  \bibinfo{pages}{11056} (\bibinfo{year}{2004}).


\bibitem[{wat({\natexlab{b}})}]{water2}
\bibinfo{note}{Unpublished results. In preparation.}

\bibitem[{\citenamefont{Berendsen et~al.}(1981)\citenamefont{Berendsen, Postma,
  van Gunsteren, and Hermans}}]{Berendsen:1981}
\bibinfo{author}{\bibfnamefont{H.~J.~C.} \bibnamefont{Berendsen}},
  \bibinfo{author}{\bibfnamefont{J.~P.~M.} \bibnamefont{Postma}},
  \bibinfo{author}{\bibfnamefont{W.~F.} \bibnamefont{van Gunsteren}},
  \bibnamefont{and} \bibinfo{author}{\bibfnamefont{J.}~\bibnamefont{Hermans}},
  in \emph{\bibinfo{booktitle}{Intermolecular Forces}}, edited by
  \bibinfo{editor}{\bibfnamefont{B.}~\bibnamefont{Pullman}}
  (\bibinfo{publisher}{Reidel}, \bibinfo{address}{Dordrecht},
  \bibinfo{year}{1981}), p. \bibinfo{pages}{331}.

\bibitem[{\citenamefont{Berendsen et~al.}(1987)\citenamefont{Berendsen,
  Grigera, and Straatsma}}]{Berendsen:1987}
\bibinfo{author}{\bibfnamefont{H.~J.~C.} \bibnamefont{Berendsen}},
  \bibinfo{author}{\bibfnamefont{J.~R.} \bibnamefont{Grigera}},
  \bibnamefont{and} \bibinfo{author}{\bibfnamefont{T.~P.}
  \bibnamefont{Straatsma}}, \bibinfo{journal}{J. Phys. Chem.}
  \textbf{\bibinfo{volume}{91}}, \bibinfo{pages}{6269} (\bibinfo{year}{1987}).

\end{thebibliography}

%
\end{document}